\documentclass[amsmath,amssymb,aps,prb,reprint,superscriptaddress,showpacs]{revtex4-1}

\usepackage{graphicx}
\usepackage{dcolumn}
\usepackage{bm}
\usepackage{amsmath}
\usepackage{epstopdf}
\usepackage{color}
\usepackage{soul}

\begin{document}


\title{Crossover from nematic to magnetic low-temperature ground state in Fe(Se,Te) compounds}

\author{Y.A. Ovchenkov}
\email[]{ovtchenkov@mig.phys.msu.ru}
\affiliation{Faculty of Physics, M.V. Lomonosov Moscow State University, Moscow 119991, Russia}
\author{D.A.~Chareev}
\affiliation{Institute of Experimental Mineralogy, RAS, Chernogolovka, 123456, Russia}
\affiliation{Ural Federal University, Ekaterinburg, 620002, Russia }
\affiliation{National University of Science and Technology `MISiS' 119049 Moscow, Russia }

\author{D.E. Presnov}
\affiliation{Skobeltsyn Institute of Nuclear Physics, Moscow 119991, Russia}
\author{O.S. Volkova}
\affiliation{Faculty of Physics, M.V. Lomonosov Moscow State University, Moscow 119991, Russia}
\affiliation{Ural Federal University, Ekaterinburg, 620002, Russia }
\affiliation{National University of Science and Technology `MISiS' 119049 Moscow, Russia }
\author{A.N. Vasiliev}
\affiliation{Faculty of Physics, M.V. Lomonosov Moscow State University, Moscow 119991, Russia}
\affiliation{National University of Science and Technology `MISiS' 119049 Moscow, Russia }
\affiliation{National Research South Ural State University, 454080 Chelyabinsk, Russia }

%

%

%
%
%
%
%
%


\begin{abstract}
A comparative analysis of the properties of FeSe${}_{1-x}$Te${}_{x}$ crystals in the range of x values of about 0.4 and pure FeSe crystals is presented. We found that the anomaly in R (T) at the structural transition for the former differs significantly from the corresponding anomaly for the latter. This indicates a change in the type of the ground state in the studied compounds. Within the framework of the crystal field model, this can be explained as a consequence of a change in the distortion of the tetrahedral environment of iron, which leads to a change in the positions of the energy levels within $t_{2g}$ multiplet. Depending on the mutual position of the degenerate xz and yz levels and the xy level, the type of transition can change from orbital ordering to magnetic ordering.
\end{abstract}

\maketitle


\section{Introduction}

\begin{figure}[h]
\centering
  \includegraphics[scale=0.5]{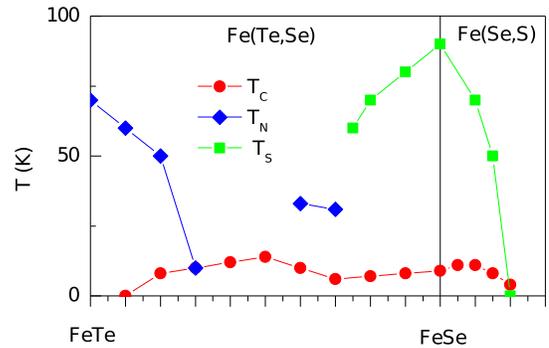}
  \caption{ Substitution-temperature phase diagram of FeSe${}_{1-x}$Te${}_{x}$ and FeSe${}_{1-x}$S${}_{x}$.}
  \label{fgr:fig1}
\end{figure}

The electronic and magnetic instability of the tetragonal plane of iron atoms surrounded by pnictides or chalcogenides is an important factor \cite{PhysRevB.79.220510, mazin2010S} in the onset of superconductivity in iron-based compounds (IBS). The phase diagram of the 11 series of IBS is very rich \cite{hsu2008superconductivity, 2017_Coldea, Bohmer2018, zhuang2014, PhysRevB.100.224516, mukasa2021high}. Fig.\ref{fgr:fig1} shows the main details of this phase diagram at ambient pressure.

The binary compositions in this series, FeSe and FeTe have phase transitions of different types. For FeTe, the structural transition around 70~K is accompanied by antiferromagnetic ordering and at low temperatures FeTe remains non-superconducting. In the case of FeSe no magnetic ordering occurs below the structural transition at around 90~K and at low temperatures FeSe is superconducting. It should be noted that the nature of the structural transition in these materials is still under debate. Any substitutions at chalcogen positions in FeSe and FeTe suppress structural transitions. For FeTe, the suppression of the magnetic transition is accompanied by the onset of superconductivity, which is very similar to the scenario for the onset of superconductivity in cuprate superconductors and suggests related mechanisms of high-temperature superconductivity.

For FeSe${}_{1-x}$S${}_{x}$ materials, with increasing x the value of T${}_{C}$ first increases slightly, and then rapidly decreases. The structural transition in this series is completely suppressed for compositions with $x$ about 0.2 \cite{Ovch_JLTP, Hosoi2016}. In this series, only crystals with $x$ lower than 0.25 have been grown \cite{Char_FeSeS_CEC}. For  FeSe${}_{1-x}$Te${}_{x}$ materials, there is a local minimum of T${}_{C}$ for compositions with a low tellurium content and a global maximum of T${}_{C}$ at $x\approx0.5$. In the high-quality crystals of FeSe${}_{0.5}$Te${}_{0.5}$ there are no structural or magnetic transitions, although with an excess of iron, magnetism in FeSe${}_{1-x}$Te${}_{x}$ can be retained \cite{OVC_PhC}.

An excess of iron in Fe${}_{1+\delta}$Se${}_{1-x}$Te${}_{x}$ crystals significantly changes their properties \cite{sun2015evolution}. Recent progress in obtaining high-quality crystals of compositions with a low tellurium content  made it possible to obtain data on the properties of compositions with a low tellurium content.

Here we present the results of studies of the transport properties of two compositions  FeSe${}_{1-x}$Te${}_{x}$  with a tellurium content of about 30-40\%. For these compositions, the change in the electronic properties during the structural transition is intrinsically different from that of pure FeSe. It may indicate that the low-temperature ground state has changed from the orbital order to the magnetic order.

\section{Experiment}

The FeSe and FeSe${}_{1-x}$Te${}_{x}$ crystals were grown using recrystallization in AlCl3/NaCl/KCl flux technique with a constant temperature gradient along the quartz ampoule \cite{CrystEngComm12.1989}. A driving force of the recrystallization process is the temperature gradient. For FeSe the temperature of the hot end of the ampoule was 430 $^{\circ}$C and the temperature of the cold end was 360 $^{\circ}$C. The synthesis was carried out for about 55 days. FeSe${}_{1-x}$Te${}_{x}$ crystals were grown in 70 days at 585 $^{\circ}$C and 495 $^{\circ}$C. The chemical composition of the crystals was determined using a Tescan Vega II XMU scanning electron microscope equipped with an INCA Energy 450 energy-dispersive spectrometer; the accelerating voltage was 20~kV. For the batch designated further as FeSe${}_{0.7}$Te${}_{0.3}$, the experimentally determined chemical composition values of Fe${}_{1+\delta}$Se${}_{1-x}$Te${}_{x}$ are $\delta{}\approx$-0.02 and $0.3>x>0.23$. For the batch designated further as FeSe${}_{0.6}$Te${}_{0.4}$ the experimental values are $\delta{}\approx$-0.02 and $0.42>x>0.33$.
Electrical measurements were done on cleaved samples with  Au/Ti or Pt contacts  made by sputtering.

\section{Results and discussion}

\begin{figure}[h]
\centering
  \includegraphics[scale=0.25]{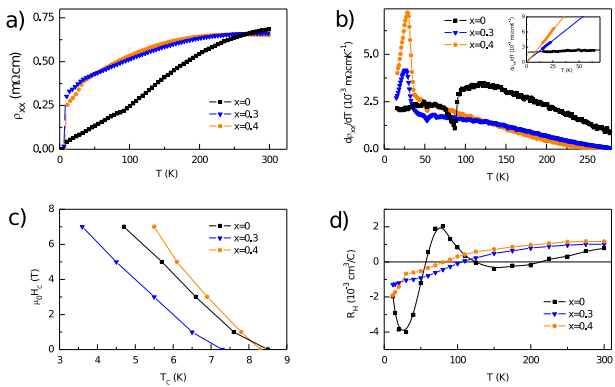}
  \caption{The properties of FeSe${}_{1-x}$Te${}_{x}$ crystals with $x$=0, 0.3, and 0.4. (a)~Temperature dependence of the resistivity $\rho_{xx}$. (b)~~Temperature dependence of the derivative of the resistivity  $d\rho_{xx}/dT$. The inset shows the linear fit for  $d\rho_{xx}/dT$(T) below the structural transition temperature. (c)~Temperature dependence of the resistive upper critical field for $H//c$. (d)~Temperature dependence of the Hall constant $R_{H}$. }
  \label{fgr:fig2}
\end{figure}

The temperature dependencies of the resistivity $\rho_{xx}$ of compositions with 30-40\% tellurium and pure FeSe differ significantly (see Fig.\ref{fgr:fig2}a)). At temperatures above the structural transition, the resistivity of compositions with tellurium depends rather weakly on temperature, which corresponds to the bad-metal state. At the same time, the temperature dependence of the derivative of resistivity $d\rho_{xx}/dT$ (Fig.\ref{fgr:fig2}b)) have no qualitative difference between the FeSe and compositions with tellurium. These dependencies in both cases are linear functions with a moderate negative slope. A negative slope may be due to the presence of an activation contribution to the resistance, which for FeSe was previously demonstrated at high temperatures \cite{sust_28_10_105009}.

On the contrary, at low temperatures, the shape of the $d\rho_{xx}/dT$(T) curves for the compositions with tellurium and FeSe is qualitatively different. In our opinion, this can indicate a change in the ground state at low temperatures.

Fig.\ref{fgr:fig2}c) shows the temperature dependence of  the resistive upper critical field for $H//c$. For pure FeSe and the studied compositions with tellurium, the critical temperatures and upper critical fields have similar values. It may be because the temperature of the superconducting transition reaches the local minimum for FeSe${}_{1-x}$Te${}_{x}$ at intermediate tellurium concentrations \cite{mukasa2021high, ovchenkov2019nematic}.

\begin{figure}[h]
\centering
  \includegraphics[scale=0.25]{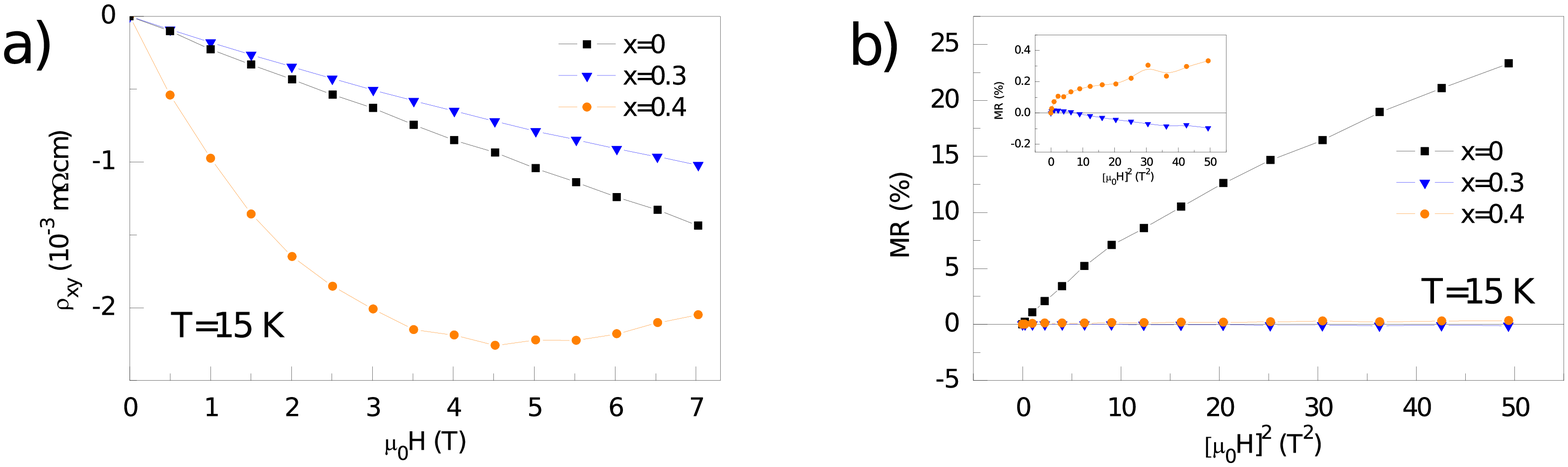}
  \caption{The properties of FeSe${}_{1-x}$Te${}_{x}$ crystals with $x$=0, 0.3, and 0.4 at 15~K. (a)~Magnetic field dependence of the Hall resistivity $\rho_{xy}$. (b)~Magnetoresistance $MR$=($\rho_{xx}$(B)-$\rho_{xx}$(0))/$\rho_{xx}$(0) versus $B^{2}$. The inset shows the zoom-in view at low $MR$.}
  \label{fgr:fig3}
\end{figure}

The temperature dependence of the Hall constant $R_{H}$ is shown in Fig.\ref{fgr:fig2}d). For FeSe${}_{1-x}$Te${}_{x}$, there is only one inversion point on $R_{H}$(T). At temperatures of structural transitions, curves $R_{H}$(T) have kinks.
Fig.\ref{fgr:fig3} shows the field dependence of the Hall component of the resistivity $\rho_{xy}$  and the transverse magnetoresistance $MR$=($\rho_{xx}$(B)-$\rho_{xx}$(0))/$\rho_{xx}$(0) versus $B^{2}$. The magnetoresistance of compositions with tellurium decreases significantly in comparison with the magnetoresistance of FeSe. This indicates a significant decrease in mobility in compositions with tellurium. Against the background of low values of magnetoresistance, a particular case of negative magnetoresistance of the composition $x=0.3$ may be due to the presence of magnetic impurities. Knowing the slope of the $MR$($B^{2}$) curve and the conductivity, the carrier mobility for FeSe${}_{0.6}$Te${}_{0.4}$ can be estimated \cite{2017ovchenkovMISM} at about 60 $cm^{2}/Vs$, and the carrier concentration - at about 4$\times$10$^{20}$ $cm^{-3}$.

\begin{table*}[ht]
\small
  \caption{ Specification of the studied FeSe${}_{1-x}$Te${}_{x}$ samples. The $T_{c}$ value is the midpoint in the R(T) superconducting transition. $T_{s}$ is the point with the maximum of absolute value of the second derivative of $\rho_{xx}$. $C_{1}$T+$C_{0}$ is the best linear fit for $d\rho_{xx}/dT$(T) in the specified temperature range.  }
  \label{tbl:T1}
\begin{ruledtabular}
  \begin{tabular}[t]{lccccc}
    Sample & $T_{C}$ (K)  & $T_{S}$ (K)  & \multicolumn{3}{c}{Linear fit for $d\rho_{xx}/dT$} \\
    &&& Range & $C_{1}$ ( $\Omega$cmK$^{-2}$) &  $C_{0}$ ( $\Omega$cmK$^{-1}$) \\
    \hline \\ [-4pt]
    FeSe  &8.5  & 90 & 15~K - 65~K & 6.4$\times$10$^{-9}$ & 2.0$\times$10$^{-6}$  \\
    FeSe${}_{0.7}$Te${}_{0.3}$  &7.3  & 31 & 15~K - 23~K & 1.6$\times$10$^{-7}$ & 1.7$\times$10$^{-7}$   \\
    FeSe${}_{0.6}$Te${}_{0.4}$ &8.3  & 33 & 15~K - 26~K & 2.6$\times$10$^{-7}$ & 2.3$\times$10$^{-7}$   \\

  \end{tabular}
\end{ruledtabular}
\end{table*}

As mentioned above, the behavior of resistance near the structural transition and below for FeSe${}_{1-x}$Te${}_{x}$ $x\ge0.3$ is significantly different from that of FeSe. First, the form of the singularity at the structural transition changes. For FeSe, when the temperature decreases, the value of the derivative $d\rho_{xx}/dT$ drops sharply and is then restored approximately to the original value. This roughly corresponds to the formation of a small step on the temperature dependence of the resistivity $\rho_{xx}$(T). Similar shape have anomalies on curves $\rho_{xx}$(T) for FeSe${}_{1-x}$S${}_{x}$ \cite{Ovch_JLTP} and FeSe${}_{1-x}$Te${}_{x}$ with a low tellurium content \cite{ovchenkov2019nematic}. At a concentration of tellurium of about 30\%, the form of the singularity changes \cite{PhysRevB.100.224516} and a sharp kink appears instead of a step. On the derivative chart, it looks like a sharp rise in the derivative $d\rho_{xx}/dT$ at the transition (see Fig.\ref{fgr:fig2}b) ).

Formally, one could argue that for substituted compositions the singularity at the transition is similar to the left half of the singularity for FeSe - the transition from the flat region of $\rho_{xx}$(T) to the low-temperature part. Therefore a change in the shape of the singularity could be caused by a change in the slope of $\rho_{xx}$(T) in the high-temperature part due to the transition to the bad-metal state. This description of the evolution of the shape of singularity during a structural transition may be correct. Nevertheless, a noticeable change in the behavior of the resistance in the region between the superconducting and structural transitions also deserves attention.

The inset in  Fig.\ref{fgr:fig2}b) shows the temperature dependence of the derivative of the resistivity $d\rho_{xx}/dT$ in the temperature range between the superconducting and structural transitions. For the studied FeSe${}_{1-x}$Te${}_{x}$ compositions, these dependencies are near-linear functions. The coefficients of linear regression of $d\rho_{xx}/dT$(T) are given in Table \ref{tbl:T1}. It can be seen that for the substituted compositions the derivative is described with good accuracy by the linear function $aT$, and for the FeSe composition the derivative of the resistance is a constant. Therefore, the type of temperature dependence of resistance changes from linear for pure FeSe to quadratic $aT^{2}$ for compositions FeSe${}_{1-x}$Te${}_{x}$ with $x$ = 0.3 and 0.4. The change in the power of the $\rho_{xx}$(T) function may indicate a change in the type of the ground state.

From the phase diagram shown in Fig.\ref{fgr:fig1}, it can be seen that the temperature of the structural transition in the 11-type iron-based superconductors reaches its local maximums in pure FeSe and FeTe. In our opinion, this may be due to a lowering of the local symmetry in the substituted compositions. In all three cases shown on the diagram, the structural transition is suppressed at a substitution level close to 25\%, that roughly corresponds to the substitution of one of the four chalcogens near each iron atom. Thus this substitution level can remove the degeneracy of the energy levels caused by the tetragonal symmetry and makes symmetry lowering in the structural transition redundant.

If we suppose that the two domes shape of the phase diagram is the result of the modulation of excess electron energy by disorder, then there can be an immediate boundary between two types of  the low-temperature order which exist in FeTe and FeSe. In the vicinity of this quantum phase transition, various anomalies in the properties of the compounds should be observed. For FeSe${}_{1-x}$Te${}_{x}$, significant changes in the properties occur at $x<0.4$. In this range of compositions, the piezoresistivity coefficient changes the sign, and the type of the majority carriers changes, which may also be connected with changes in the crystal field \cite{ovch_sust_21}.

\begin{figure}[h]
\centering
  \includegraphics[scale=0.4]{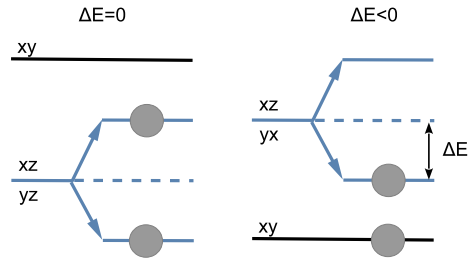}
  \caption{Conventional scheme of electronic instability of the Jahn-Teller type depending on the distortion in the tetrahedral environment of Fe.}
  \label{fgr:fig4}
\end{figure}

The crystal field can influence the type of ordering. In the studied compounds, the dispersion for $d$-orbitals is several electron volts. Nevertheless, the crystal field determines which orbitals are at the Fermi level. Under substitution, the lattice parameters change anisotropically. In this case, the distortion of the tetragonal environment of iron changes, which should cause changes in the relative position of the levels in the $t_{2g}$ multiplet. Depending on the degree of the population of the states corresponding to different orbitals, either conditions for orbital ordering ( see Fig.\ref{fgr:fig4}) or for magnetic ordering can be realized.

\section{Conclusion}

The study of the electron transport properties in FeSe${}_{1-x}$Te${}_{x}$ compositions with $x$=0, 0.3, and 0.4 suggests the crossover in the ground state from nematic orbital order in compositions with a low tellurium content to magnetic order in compositions with $x$ greater than about 0.3. If confirmed, this can be an example of the itinerant system with magnetic and orbital order competition, that deserves further in-depth study.

\section{Acknowledgments}

This work has been supported by Russian Foundation for Basic Research through Grant 20-02-00561A and by Interdisciplinary Scientific and Educational School of Moscow State University “Photonic and Quantum Technologies. Digital Medicine”, the research infrastructure of the “Educational and Methodical Center of Lithography and Microscopy”, M.V. Lomonosov Moscow State University was used . OSV acknowledge the financial support by Russian Science Foundation through the project 22-42-08002. ANV acknowledge the financial support by the Megagrant 075-15-2021-604 of the Government of Russian Federation.

\section*{References}
 \bibliographystyle{unsrt} 
\bibliography{FeSeTe_mag.bib}

\begin{thebibliography}{10}

\bibitem{PhysRevB.79.220510}
M.~D. Johannes and I.~I. Mazin.
\newblock Microscopic origin of magnetism and magnetic interactions in
  ferropnictides.
\newblock {\em Phys. Rev. B}, 79:220510, Jun 2009.

\bibitem{mazin2010S}
I.~I. Mazin.
\newblock Superconductivity gets an iron boost.
\newblock {\em Nature}, 464(7286):183--186, 2010.

\bibitem{hsu2008superconductivity}
Fong-Chi Hsu, Jiu-Yong Luo, Kuo-Wei Yeh, Ta-Kun Chen, Tzu-Wen Huang, Phillip~M
  Wu, Yong-Chi Lee, Yi-Lin Huang, Yan-Yi Chu, Der-Chung Yan, et~al.
\newblock Superconductivity in the {PbO-type} structure {$\alpha$-FeSe}.
\newblock {\em Proceedings of the National Academy of Sciences},
  105(38):14262--14264, 2008.

\bibitem{2017_Coldea}
A.~I. Coldea and M.~D. Watson.
\newblock The key ingredients of the electronic structure of {FeSe}.
\newblock {\em Annual Review of Condensed Matter Physics}, 9(1):125--146, 2018.

\bibitem{Bohmer2018}
A.~E. Bohmer and A.~Kreisel.
\newblock {Nematicity, magnetism and superconductivity in {FeSe}}.
\newblock {\em {Journal of Physics-Condensed Matter}}, {30}({2}), {JAN 17}
  {2018}.

\bibitem{zhuang2014}
J.~Zhuang, W.~K. Yeoh, X.~Cui, X.~Xu, Y.~Du, Z.~Shi, S.~P. Ringer, X.~Wang, and
  S.~X. Dou.
\newblock Unabridged phase diagram for single-phased {FeTe$_{1-x}$Se$_{x}$}
  thin films.
\newblock {\em Scientific reports}, 4:7273, 2014.

\bibitem{PhysRevB.100.224516}
Kotaro Terao, Takanari Kashiwagi, Tomoyuki Shizu, Richard~A. Klemm, and Kazuo
  Kadowaki.
\newblock Superconducting and tetragonal-to-orthorhombic transitions in single
  crystals of {${\mathrm{FeSe}}_{1\ensuremath{-}x}{\mathrm{Te}}_{x}$
  $(0\ensuremath{\le}$ $x$ $\ensuremath{\le}0.61)$}.
\newblock {\em Phys. Rev. B}, 100:224516, Dec 2019.

\bibitem{mukasa2021high}
K~Mukasa, K~Matsuura, M~Qiu, M~Saito, Y~Sugimura, K~Ishida, M~Otani, Y~Onishi,
  Y~Mizukami, K~Hashimoto, et~al.
\newblock High-pressure phase diagrams of {FeSe$_{1-x}$Te$_{x}$}: correlation
  between suppressed nematicity and enhanced superconductivity.
\newblock {\em Nature communications}, 12(1):1--7, 2021.

\bibitem{Ovch_JLTP}
Y.~A. Ovchenkov, D.~A. Chareev, D.~E. Presnov, O.~S. Volkova, and A.~N.
  Vasiliev.
\newblock Superconducting properties of {FeSe$_{1-x}$S$_{x}$} crystals for x up
  to 0.19.
\newblock {\em J. Low Temp. Phys.}, 185:467--473, 2016.

\bibitem{Hosoi2016}
S.~Hosoi, K.~Matsuura, K.~Ishida, H.~Wang, Y.~Mizukami, T.~Watashige,
  S.~Kasahara, Y.~Matsuda, and T.~Shibauchi.
\newblock Nematic quantum critical point without magnetism in
  {FeSe$_{1-x}$S$_{x}$} superconductors.
\newblock {\em Proceedings of the National Academy of Sciences},
  113(29):8139--8143, 2016.

\bibitem{Char_FeSeS_CEC}
D.~A. Chareev, Y.~A. Ovchenkov, L.~V. Shvanskaya, A.~M. Kovalskii,
  M.~Abdel-Hafiez, D.~Traine, E.~Lechner, M.~Iavarone, O.~S. Volkova, and A.~N.
  Vasiliev.
\newblock Single crystal growth{,} transport and scanning tunneling microscopy
  and spectroscopy of {FeSe$_{1-x}$S$_{x}$}.
\newblock {\em CrystEngComm}, 20:2449--2454, 2018.

\bibitem{OVC_PhC}
Y.A. Ovchenkov, D.A. Chareev, E.S. Kozlyakova, O.S. Volkova, and A.N. Vasiliev.
\newblock Coexistence of superconductivity and magnetism in
  {Fe$_{1+\delta}$Te$_{1-x}$Se$_{x}$} (x=0.1, 0.2, 0.28, 0.4 and 0.45).
\newblock {\em Physica C: Superconductivity}, 489:32--35, 2013.

\bibitem{sun2015evolution}
Yue Sun, Toshihiro Taen, Tatsuhiro Yamada, Yuji Tsuchiya, Sunseng Pyon, and
  Tsuyoshi Tamegai.
\newblock Evolution of superconducting and transport properties in annealed
  {FeTe$_{1-x}$Se$_{x}$} (0.1$\leq$x$\leq$0.4) multiband superconductors.
\newblock {\em Superconductor Science and Technology}, 28(4):044002, 2015.

\bibitem{CrystEngComm12.1989}
D.~Chareev, E.~Osadchii, T.~Kuzmicheva, J.-Y. Lin, S.~Kuzmichev, O.~Volkova,
  and A.~Vasiliev.
\newblock Single crystal growth and characterization of tetragonal
  {FeSe$_{1-x}$S$_{x}$} superconductors.
\newblock {\em CrystEngComm}, 15:1989--1993, 2013.

\bibitem{sust_28_10_105009}
S.~Karlsson, P.~Strobel, A.~Sulpice, C.~Marcenat, M.~Legendre, F.~Gay,
  S.~Pairis, O.~Leynaud, and P.~Toulemonde.
\newblock Study of high-quality superconducting {FeSe} single crystals:
  crossover in electronic transport from a metallic to an activated regime
  above 350 k.
\newblock {\em Superconductor Science and Technology}, 28(10):105009, 2015.

\bibitem{ovchenkov2019nematic}
Y.~A. Ovchenkov, D.~A. Chareev, V.~A. Kulbachinskii, V.~G. Kytin, D.~E.
  Presnov, Y.~Skourski, L.~V. Shvanskaya, O.~S. Volkova, D.~V. Efremov, and
  A.~N. Vasiliev.
\newblock Nematic properties of {FeSe$_{1-x}$Te$_{x}$} crystals with a low {Te}
  content.
\newblock {\em arXiv1909.00711}, 2019.

\bibitem{2017ovchenkovMISM}
Y.A. Ovchenkov, D.A. Chareev, V.A. Kulbachinskii, V.G. Kytin, D.E. Presnov,
  Y.~Skourski, O.S. Volkova, and A.N. Vasiliev.
\newblock Magnetotransport properties of {FeSe} in fields up to 50{T}.
\newblock {\em Journal of Magnetism and Magnetic Materials}, 459:221 -- 225,
  2018.

\bibitem{ovch_sust_21}
Y.~A. Ovchenkov, D.~Chareev, E.~S. Kozlyakova, E.~Levin, M.~G. Mikheev, D.~E.
  Presnov, A.S. Trifonov, O.~Volkova, and A.~Vasiliev.
\newblock {Phase separation near the charge neutrality point in FeSe1-xTex
  crystals with x $< $0.15}.
\newblock {\em Superconductor Science and Technology}, 2021.

\end{thebibliography}



\end{document}